\documentstyle[multicol,epsf,aps]{revtex}
\begin{document}
\title{Shock-Like Dynamics of Inelastic Gases}
\author{E.~Ben-Naim$\dag$, S.~Y.~Chen$\dag$, G.~D.~Doolen$\dag$, 
and S.~Redner$\ddag$}
\address{$\dag$Theoretical Division and Center for Nonlinear Studies,  
Los Alamos National Laboratory, Los Alamos, NM 87545}
\address{$\ddag$Center for Polymer Studies and Department of Physics,
Boston University, Boston, MA 02215}
\maketitle
\begin{abstract}
  We provide a simple physical picture which suggests that the
  asymptotic dynamics of inelastic gases in one dimension is
  independent of the degree of inelasticity.  Statistical
  characteristics, including velocity fluctuations and the velocity
  distribution are identical to those of a perfectly inelastic sticky
  gas, which in turn is described by the inviscid Burgers equation.
  Asymptotic predictions of this continuum theory, including the $t^{-2/3}$
  temperature decay and the development of discontinuities in the
  velocity profile, are verified numerically for inelastic gases.
\end{abstract}
\begin{multicols}{2} 
  Gases of inelastically colliding particles model the dynamics of
  granular materials \cite{pkh,lpk}, geophysical flows \cite{csc}, and
  large-scale structure of matter in the universe \cite{sz}.
  Typically, a fraction of the kinetic energy is dissipated in each
  collision, leading to interparticle velocity
  correlations, a clustering instability \cite{gz,dlk,gzb,kwg,sg}, and in
  the absence of external energy input, an inelastic collapse
  \cite{bm,my,db,ernst,vn,brey}. The last feature presents an obstacle
  to long-time simulations, as an infinite number of collisions occur
  in a finite time.
  
  In this Letter, we propose that a freely evolving inelastic gas is
  asymptotically in the universality class of a completely inelastic,
  sticky gas.  Specifically, the temperature decreases in time as $t^{-2}$
  over an intermediate range, but asymptotically decays as $t^{-2/3}$.
  To test this hypothesis, we employ a simulation in which collisions
  between particles with sufficiently small relative velocities are
  perfectly elastic.  This method allows us to bypass the inelastic
  collapse and probe the asymptotic regime.
  
  We consider $N$ identical point particles undergoing inelastic
  collisions in a one-dimensional periodic system of length $L$.  The
  particles have typical interparticle spacing $x_0=L/N$ and their
  typical velocity is $v_0$.  We employ dimensionless space and time
  variables, $x\to x/x_0$, and $t\to t v_0/x_0$, thereby rescaling the
  ring length to $N$.  Inelastic and momentum conserving collisions
  are implemented by changing the sign of the relative velocity and
  reducing its magnitude by a factor $r=1-2\epsilon$, with $0\leq
  r\leq 1$, after each collision.  It is convenient to view the
  particle identities as ``exchanged'' upon collision, so that in a
  perfectly elastic collision the particles merely pass through each
  other, while for a small inelasticity each particle suffers a small
  deflection.  The outcome of a collision between a particle with
  velocity $v$ and another particle with velocity $u$ is therefore
\begin{equation} 
\label{rule}
v\to v-\epsilon(v-u).
\end{equation}

The granular temperature, or velocity fluctuation, 
$T(t)=\langle v^2(t)\rangle-\langle v(t)\rangle^2$, can be estimated
in the intermediate time regime by considering the outcome of a single
collision under the assumption that the system remains homogeneous.
In each such collision, the energy lost is $\Delta
T\propto-\epsilon(\Delta v)^2$, with $\Delta v$ the relative
velocity, while the time between collisions is $\ell/\Delta v$.
Assuming homogeneity, we neglect fluctuations in the mean-free path
$\ell\cong 1$ and posit a single velocity scale so that $v\sim \Delta
v\sim T^{1/2}$.  The temperature therefore obeys the rate equation
$dT/dt\propto-\epsilon\, T^{3/2}$ giving 
\begin{equation}
\label{t2} 
T(t)\sim (1+A\epsilon t)^{-2},
\end{equation}
with $A$ a constant of order unity \cite{pkh}.  For small times $t\ll
t_{\rm dissip}\sim \epsilon^{-1}$ dissipation is negligible and the
temperature does not evolve -- the gas is effectively
elastic.  For larger times, the dissipation leads to a $\epsilon^{-2}
t^{-2}$ temperature decay.

However, this behavior {\em cannot\/} be valid asymptotically, as the
temperature must decrease monotonically with increasing dissipation.
Moreover, the temperature is bounded from below by that of the
perfectly inelastic gas with a vanishing restitution coefficient,
$r=0$.  For such a sticky gas, the temperature decays as $t^{-2/3}$
and the typical cluster mass grows as $t^{2/3}$ \cite{cpy}.  This
behavior is reminiscent of diffusion-controlled two species
annihilation, where a small reaction probability results in a
homogeneous intermediate time regime where the density follows a
$t^{-1}$ mean-field decay, even for low spatial dimension $d$.
However, at long times single-species domains which are opaque to
opposite-species particles form and a slower $t^{-d/4}$ density decay
follows\cite{kr}.

For the inelastic gas, we argue that the role of the reaction
probability is played by $\epsilon$.  For small $\epsilon$, a particle
can penetrate through a domain of $N<N_c(\epsilon)\sim \epsilon^{-1}$
coherently-moving particles without experiencing a substantial
deflection.  The critical cluster size $N_c(\epsilon)$ may be
estimated by considering a collision between a moving particle and a
cluster of $N$ stationary particles.  From Eq.~(\ref{rule}), each
collision between the incident particle and the next particle in the
cluster reduces the incident particle velocity by roughly $\epsilon$.
After $N$ collisions the incident velocity is $v_N\approx 1-N\epsilon$.  For
the particle to pass through the cluster, the number of particles must
therefore be less than $\epsilon^{-1}$.  It is in this range of
cluster sizes that the system remains spatially homogeneous and the
mean-field decay $T\sim\epsilon^{-2}t^{-2}$ holds.

However, once the cluster size is larger than $N_c(\epsilon)\sim
\epsilon^{-1}$, an incident particle is ``absorbed'' and the decay
follows that of the perfectly inelastic gas.  That is, domains larger
than $N_c(\epsilon)$ are opaque and present an effective restitution
coefficient $r_{\rm eff}\equiv 0$ to incident particles.  We argue 
that a similar sticking mechanism also governs cluster-cluster 
collisions.  The crossover time $t_{\rm stick}$ between these two
regimes is obtained by matching the intermediate and long-time
temperature decays, $\epsilon^{-2} t^{-2}$ and $t^{-2/3}$, to give
$t_{\rm stick}\sim\epsilon^{-3/2}$.

These arguments suggest the temperature decay
\begin{equation}
\label{t223}
T(t)\sim \cases{1&$t\ll \epsilon^{-1}\equiv t_{\rm dissip}$;\cr 
\epsilon^{-2}t^{-2}&$\epsilon^{-1}\ll t\ll 
\epsilon^{-3/2}\equiv t_{\rm stick}$;\cr 
t^{-2/3}&$\epsilon^{-3/2}\ll t\ll N^{3/2}$;\cr 
N^{-1}&$N^{3/2}\ll t$.}
\end{equation}
The last regime reflects the final state of a finite $N$-particle
system, namely a single cluster of mass $m=N$, velocity $v\sim
N^{-1/2}$, and therefore energy $T\sim v^2\sim N^{-1}$.  This final
velocity follows from momentum conservation in which the total momentum $P$ is
the sum of $N$ individual random momenta of order unity.
Consequently, $P\propto N^{1/2}$ and $v=P/m\sim N^{-1/2}$. 

The above crossover picture applies equally well to moderately
inelastic gases where both $t_{\rm dissip}$ and $t_{\rm stick}$ are of
order unity and the asymptotic behavior is realized immediately.
While weakly inelastic systems with a small number of particles
$N<\epsilon^{-1}$ will avoid the clustering regime and follow the
$t^{-2}$ cooling law indefinitely, the $t^{-2/3}$ sticky gas regime is
always reached in the thermodynamics limit, $N\to\infty$. Therefore,
the $t^{-d/2}$ decay conjectured in \cite{be,cdnt} based on two- and
three-dimensional simulations does not extend to lower dimensions.

To probe the long-time behavior, we performed numerical simulations of
$N$ particles which are initially equally spaced ($\Delta x=1$) and
uniformly distributed (on $[-1,1]$) in velocity.  We implemented an
event-driven simulation, keeping the collision times always sorted to
facilitate identification of the next event.  To circumvent the
inelastic collapse, elastic collisions were implemented whenever the
relative velocity of the colliding particles fell below a
pre-specified threshold, $\Delta v<\delta$ \cite{csc,lm}. In fact, the
restitution coefficient for deformable spheres does approach unity
when $\Delta v\to 0$ as a consequence of the nonlinear Hertz contact
law \cite{bp}.

Fig.~1 shows that the temperature of the freely cooling inelastic gas
asymptotically decays as $t^{-2/3}$, {\em independent\/} of the
restitution coefficient.  Moreover, the time scale over which this
decay occurs diverges in the limit of vanishing dissipation.  We also
simulated the completely inelastic gas ($r=0$) where particles
aggregate (and conserve momentum) upon collision.  This gives an
identical asymptotic temperature decay as the partially inelastic gas.
From Fig.~1, notice that the two crossover times $t_{\rm
  dissip}\sim\epsilon^{-1}$ and $t_{\rm stick}\sim\epsilon^{-3/2}$ are
consistent with the data for the cases $r=0.9$ and $0.99$, and that
the temperature is of the appropriate order $T(t_{\rm
  stick})\sim\epsilon$ at the homogeneous-sticky crossover.  On the
other hand for $r=0.5$, the intermediate $t^{-2}$ regime no longer
exists and only the sticky gas behavior is realized.
\begin{figure}
\narrowtext
\centerline{\epsfxsize=7.6cm\epsfbox{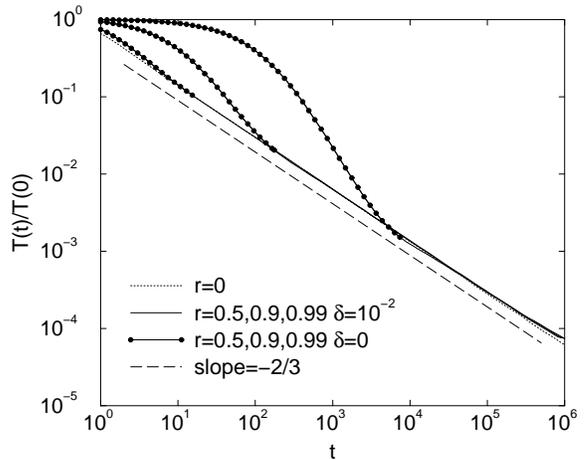}}
\caption{Temperature $T(t)$ 
  versus time for the freely cooling inelastic gas with restitution
  coefficients $r=0.5$, 0.9, and 0.99.  The data represents averages
  over 10 realizations of $N=10^6$ particles and $10^4$
  collisions per particle.  Also shown is a simulation for a
  sticky gas ($r=0$).  A dashed line of slope $-2/3$ is plotted as
  reference.  Least-square fits to the post-crossover data with velocity threshold
  $\delta=10^{-2}$ yield the decay exponents $0.67$, $0.67$, and
  $0.66$ for $r=0$, $0.5$, and $0.9$ respectively.}
\end{figure}
\vspace{-.2in}
\begin{figure}
\narrowtext
\centerline{\epsfxsize=7.6cm\epsfbox{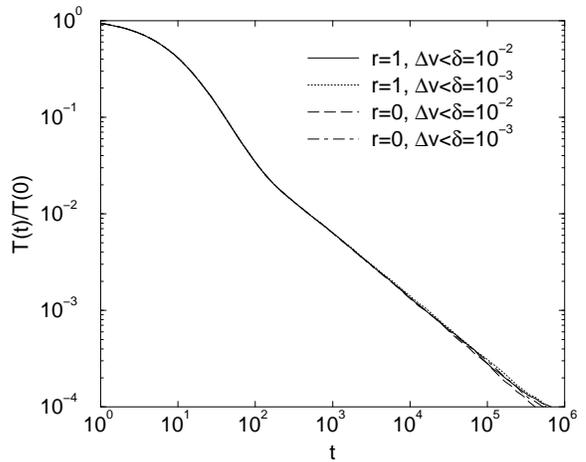}}
\caption{Role of threshold velocity and the collision mechanism.
  Temperature decay for $r=0.9$, $N=10^6$ with different threshold
  velocities $\delta=10^{-2}$, $\delta=10^{-3}$, and different
  subthreshold collision mechanisms (both sticky and elastic). }
\end{figure}
\vspace{-.05in}
The crossover between the homogeneous gas regime and the sticky gas
regime is ultimately related to the inelastic collapse. In Fig.~1,
simulation results for purely inelastic collisions are also shown
(dots); these coincide with the results for velocity threshold
$\delta=10^{-2}$.  However, as soon as the first cluster is formed,
the number of collisions becomes infinite and the simulation does not
proceed any further in time.  Indeed, the last data point for these
simulations marks the transition to the sticky gas regime.  This
provides an additional confirmation for the crossover picture, as the
typical cluster mass of the sticky gas, $m(t_{\rm stick})\sim t_{\rm
  stick}^{2/3}\sim \epsilon^{-1}$, matches the critical mass for the
appearance of opaque clusters.
       
\begin{figure}
\narrowtext
\centerline{\epsfxsize=8cm\epsfbox{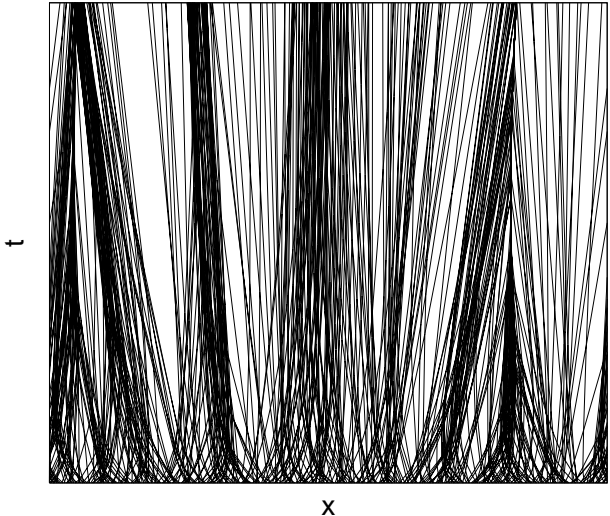}}
\vspace{0.1in}
\caption{Space-time evolution of a $500$ particle system with $r=0.9$ and 
  $\delta=10^{-2}$, up to $t=600$.}
\end{figure}

To validate the simulation method, we checked that results are
independent of the cutoff value (provided it is sufficiently
small) as well as the sub-threshold collision mechanism (Fig.~2). In
principle, the results can be trusted as long as the typical velocity
is much larger than the cutoff, $v\sim t^{-1/3}\gg \delta$, {\it
  i.e.}, up to time $t_{\rm valid}\sim\delta^{-3}$.  As shown in the
figure, the results for $\delta=10^{-2}$ and $10^{-3}$ nearly coincide
until $t=10^6$, consistent with our expectation.  Furthermore, the
space-time evolution of a weakly inelastic gas illustrates how aggregation
eventually dominates (Fig.~3).

\begin{figure}
\narrowtext
\centerline{\epsfxsize=7.6cm\epsfbox{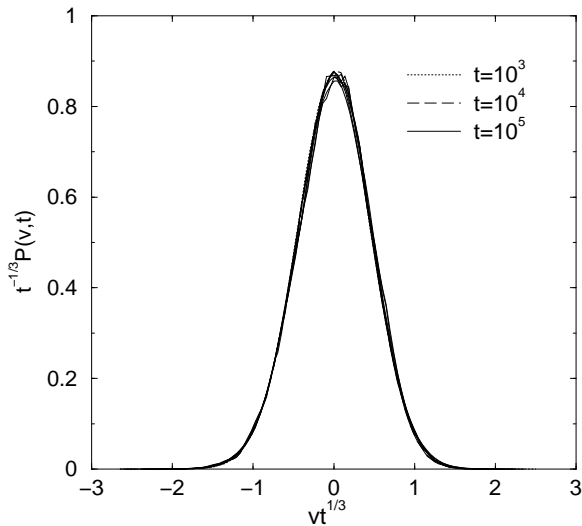}}
\caption{Scaling of the velocity distribution for restitution 
  coefficients $r=0$, $0.5$, and $0.9$, at times $t=10^3$, $10^4$,
  $10^5$.  The nine data sets represent averages over 200 realizations
  in a system of $N=10^5$ particles and cutoff $\delta=10^{-3}$.}
\end{figure}

We now investigate whether the velocity distribution, and not merely
the overall velocity scale, is also independent of $r$.  We therefore
computed this distribution for $r=0.9$ (weakly inelastic), $r=0.5$
(moderately inelastic), and $r=0$ (perfectly inelastic) at three very
different times which are well into the clustering regime. For the
$r=0$ case, the cluster velocity was weighted by the cluster mass, to
compare with the $r>0$ cases.  As shown in Fig.~4, the normalized
velocity distribution
\begin{equation}
\label{pvscl}
P(v,t)\sim t^{1/3}\Phi(vt^{1/3}),
\end{equation}
is described by an identical scaling function $\Phi(z)$ for these
widely different values of $r$.  This universality provides further
confirmation that the asymptotic behavior for any $r<1$ is governed by 
the $r=0$ ``fixed point''.

Further insights about the behavior of the inelastic gas are provided
by the connection to the Burgers equation \cite{sz,jmb}.  Since sticky
gases are described by the inviscid ($\nu\to 0$) limit of the Burgers
equation
\begin{equation} 
\label{bur}
v_t+vv_x=\nu v_{xx}, 
\end{equation}
supplemented by the continuity equation \hbox{$\rho_t+(\rho v)_x=0$},
we conclude that this continuum theory also describes the asymptotics
of the inelastic gas in the thermodynamic limit.  The Burgers equation
may be reduced to the diffusion equation by the Hopf-Cole
transformation $v=-2\nu(\ln u)_x$, and therefore is solvable.  In our
case, the relevant initial condition is delta-correlated velocities
$\langle v_0(x)v_0(x')\rangle=\delta(x-x')$.  The resulting velocity
profile is discontinuous, and the corresponding shocks can be
identified with clusters in the sticky gas.  Indeed, both shock
coalescence processes and cluster-cluster collisions in the sticky gas
conserve mass and momentum. 

The relation to the Burgers equation is useful in several ways.
First, statistical properties of the shock coalescence process have
been established analytically \cite{fm}.  For example, the tail of the
particle velocity distribution (\ref{pvscl}) is suppressed according
to
\begin{equation} 
\label{pvtail}
\Phi(z)\sim \exp(-{\rm const.}\times |z|^3\,),\qquad |z|\gg 1.
\end{equation}
This behavior can be understood by considering the density of the
fastest (order unit velocity) particles.  For such a particle to
maintain its velocity to time $t$, it must avoid collisions.  This
requires that an interval of length $\propto t$ ahead of the particle
must be initially empty \cite{bkr}.  For an initially random spatial
distribution, the probability of finding such an interval decays
exponentially with length; thus $P(1,t)\sim \exp(-{\rm const.}\times
t)$.  Using \hbox{$\Phi(z)\sim \exp(-{\rm const.}\times
  |z|^{\gamma})$} and $z=vt^{1/3}$ then yields $\gamma=3$.
Interestingly, over most of the range of scaled velocities, the
numerically obtained velocity distribution deviates only slightly from
a Gaussian, reflecting the small constant in (\ref{pvtail}) \cite{fm}.

Another important prediction of Eq.~(\ref{bur}) is that the velocity is
linear in the Eulerian coordinate $x$ and the Lagrangian coordinate
$q(x,t)$
\begin{equation} 
\label{vx}
v(x,t)={x-q(x,t)\over t}.
\end{equation}
This form also characterizes the asymptotic velocity profile of
inelastic gases.  Fig.~5 shows such a sawtooth velocity profile from
an inelastic gas simulation.  The slopes of the linear segments of the
profile are consistent with the $t^{-1}$ prediction of Eq.~(\ref{vx}).  
The inelastic collapse is simply a finite time singularity characterized 
by the development of a discontinuity in the velocity profile, {\it i.e.}, 
a shock.

\begin{figure} 
\narrowtext
\centerline{\epsfxsize=9cm \epsfbox{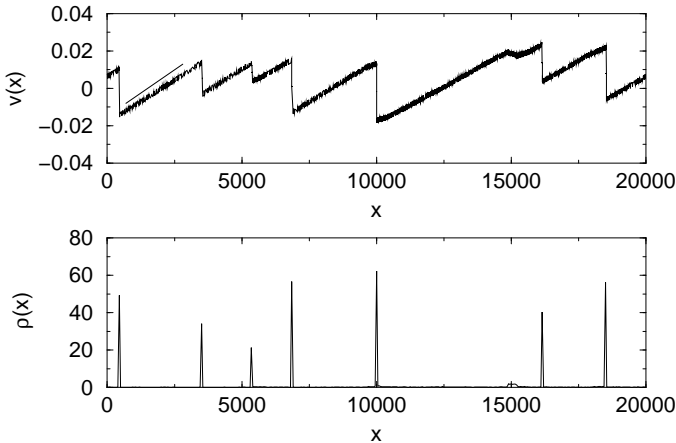}}
\caption{Shock profile of inelastic gases. 
  Density and velocity are plotted at time $t=10^5$ in a system with
  $N=2\times 10^4$ particles, $r=0.99$ and $\delta=10^{-4}$.  A line of slope
  $t^{-1}$ is plotted for reference.  The number of shocks is consistent with
  the expected number $Nt^{-2/3}\protect\cong 9$.}
\end{figure}

In higher dimensions as well, the temperature of an inelastic gas is a
monotonically increasing function of $r$ and hence, it is bounded from
below by the $r=0$ case. Therefore, we speculate that $r=0$ remains
the fixed point in higher dimensions. On the other hand, the Burgers
equation \hbox{${\bf v}_t+{\bf v}\cdot \nabla{\bf v}=\nu\nabla^2 {\bf
    v}$} approximately describes the sticky gas in the limit $\nu\to
0$ \cite{sz}.  The known $t^{-d/2}$ temperature decay of the Burgers
equation \cite{sz}, valid for $2\leq d\leq 4$ (with possible
logarithmic corrections at the crossover dimensions), then yields
\begin{equation}
T(t)\sim \cases{1&$t\ll \epsilon^{-1}$;\cr 
\epsilon^{-2}t^{-2}&$\epsilon^{-1}\ll t\ll \epsilon^{-{4/(4-d)}}$;\cr 
t^{-d/2}&$\epsilon^{-{4/(4-d)}}\ll t\ll N^{2/d}$;\cr
N^{-1}&$N^{2/d}\ll t$.} 
\end{equation}
Interestingly, both the decay exponents \cite{be,cdnt}, the formation
of string-like clusters \cite{gz,my,lm}, and even the possibility of a
percolating network of clusters \cite{jt}, features that were found
primarily numerically, are all predicted by the Burgers equation.
Additionally, the critical cluster size increases with the dimension
according to $N_c(\epsilon)\sim \epsilon^{-2d/(4-d)}$, suggesting that
the inelastic collapse is avoided when $d>d_c=4$, and that the
homogeneous gas behavior $T\sim\epsilon^{-2}t^{-2}$ holds indefinitely
above this critical dimension.

In summary, our results suggest that the asymptotic behavior of a
one-dimensional inelastic gas with many particles is governed by the
$r=0$ sticky gas fixed point, and that the appropriate continuum
theory is the inviscid Burgers equation.  This connection provides
several exact statistical properties of inelastic gases.  Conversely,
inelastic gases may provide a useful tool to study shock dynamics.
The suggestive behavior of the inelastic gas in high dimensions
deserves careful investigation.

This research is supported by the Department of Energy under contract
W-7405-ENG-36 (at LANL), and by the NSF under grant DMR9632059 (at
BU). We thank G.~P.~Berman and P.~L.~Krapivsky for useful discussions.

\end{multicols} 
\end{document}